\documentstyle[aps,prl,multicol,epsf]{revtex}
\begin{document}
\draft
\title
{Comment on:\\ Preserving Coherence in Quantum Computation by Pairing 
Quantum Bits}
\author {P. Zanardi $^{1,2}$ and M. Rasetti $^{2,3}$}
\address{$^{1}$ ISI Foundation, Villa Gualino, Torino \\
 $^2$ Unit\`a INFM, Politecnico di Torino,\\
$^3$ Dipartimento di Fisica, Politecnico di Torino\\
Corso Duca degli Abruzzi 24, I-10129 Torino, Italy
}
\maketitle
 \begin{abstract}
{We argue that several claims of paper {\sl Phys. Rev. Lett} {\bf 79}, 1953
(1997), by Lu-Ming Duan and Guang-Can Guo, are questionable. 
In particular we stress that the environmental noise
considered by the authors belongs to a very special class}
\end{abstract}
\begin{multicols}{2}[]
\narrowtext
Preserving  quantum coherence
in any computing system is an essential requirement
in order to fully exploit the new possibilities opened by quantum mechanics
in the field of information-processing \cite{HARA}.  
Up to now the most popular approach to overcome this difficulty
has been based on Quantum Error Correction (QEC) techniques that 
actively stabilize  quantum states.
These basically consist in  encoding
quantum information in a subtly redundant fashion 
in such a way that even if environmental noise is present
the uncorrupted information can be recovered by carrying on 
conditional quantum operations \cite{ERROR}.
Another strategy to prevent decoherence consists in a passive
stabilization in which one looks for error prevention rather than
error correction \cite{PALMA}. 
In reference \cite{CHINA} the authors suggest a new coding --
belonging to this latter class-- 
 based on a so-called Free Hamiltonian Elimination (FHE)
and by pairing qubits to ancillary ones in such a way that each pair
is collectively coupled with the environment.\\
In this note we would like to comment on several points
concerning the claims of the authors of \cite{CHINA}.\\
To begin with, Duan and Guo   maintain that the considered Hamiltonian 
describes the most 
general dissipation of a system of $L$ qubits coupled with some (bosonic)
environment. The interaction Hamiltonian for the $l$-th qubit 
has the form 
\begin{equation}
 H_I^l= S_l\otimes B_l, 
\label{general1}
\end{equation}
where $S_l$ is a generic {\it hermitian} one-qubit
operators ($S_l\in sl(2)$) and $B_l$ is an (hermitian) environment operator.
The total qubit-environment interaction is, of course, given by 
$H_I=\sum_l^L H_I^l.$
The above interaction is by far {\it not} the most general interaction
between a qubit and a multi-mode bosonic environment, which is instead
\begin{equation}
\tilde H_I^{l} = \sum_{k\alpha} ( \sigma^\alpha_l\otimes B_{\alpha k}^l 
+\mbox{ h.c.}),
\label{general2}
\end{equation}
where $k$ labels the environment modes,  $\alpha$ labels the generators
of $sl(2),$ and $B_{\alpha k}$ is a {\it generic} operator acting
on the Hilbert space of th $k$-th mode of the bath.
This interaction generally does not conserve {\it any} pure-qubit observable.  
To reduce (\ref{general2}) to the form (\ref{general1})
one has to make {\it two} assumptions: i) the environment is coupled
only with hermitian elements of $sl(2)$, ii)  the 
$B_{\alpha k}^l$'s do not depend on the generator label $\alpha.$\\
Actually the coupling (\ref{general1}) is nothing but a {\it pure dephasing}
coupling  inducing  suppression of the off-diagonal elements of
density matrix written in the basis of the eigenstates
of the (conserved) $S_l.$
Of course this interaction can be cast in the usual form  $\sigma_l^z\otimes B_l,$
 by a simple $SU(2)$-transformation \cite{EFF}.\\
In passing we notice also that when the ancillary qubits are introduced
the authors are not simply assuming that each pair is coupled
with the same environment modes but also that  the coupling
has the {\it same strength} for both the qubits.\\
Now we turn to consider the FHE.
It relies on the introduction of an  external driving Hamiltonian
$H_{drv},$
a procedure  necessary in order to cast the whole hamiltonian
in the form $H=\sum_l X_l\otimes B_l$  where the $X_l$'s 
\footnote{ In the notations of \cite{CHINA}, $X_l=S_l+S_{l^\prime}.$}
and the $B_l$'s  
are, respectively, qubit-pair and environment operators.
The proposed --manifestly noiseless-- 
 enconding is then given by the tensor product
of the null eigenspaces ($2$-fold degenerate) of the $X_l$'s, and
it requires therefore two physical qubits per logical qubit.
We should like to stress that -- also in the restricted class of couplings
considered by the authors -- this procedure appears quite uneffective
since the design of the external driving Hamiltonian requires full 
knowledge of the coupling constants between the environment and the
qubit-pairs. 
Such a fine-tuning of $H_{drv}$ -- even leaving aside
the obvious pratical limitations -- amounts to a knowledge of the 
noise structure which is, {\it in principle},
highly unlikely to be achieved.\\
In the last part of \cite{CHINA}
the authors address the question of gate operations.
They state (equation (10)) that the allowed gate Hamiltonians
$H_g$ have to satisfy the constraint
\begin{equation}
[H_g,\, X_l]=n_l\,{\bf I}\quad(n_l \in{\bf{C}}),
\label{comm}
\end{equation}
where $\bf I$ is the identity operator.
The above constraint implies 
 that the unitary evolution
of the $X_l$'s induced by $H_g$ is a simple $c$-number shift:
$e^{-i\,H_g\,t}\,X_l\,e^{i\,H_g\,t}= X_l-i\,t\,n_l$.
The kernel of $X_l$ is therefore mapped onto the $X_l$-eigenspace
with eigenvalue $i\,t\,n_l.$
From this follows that the gate operation does not spoil
the coherence-preserving properties of the code.\\
It is an elementary matter to show that in a finite-dimensional
Hilbert space  relation (\ref{comm}) is {\it a priori} fulfilled
only for $n_l=0.$
Indeed it is from the hermiticity of $H_g$ and $X_l$ that 
the requirement 
that the r.h.s of (\ref{comm}) has to be anti-hermitian
-- hence pure immaginary in the $c$-number case considered here -- follows.
On the other hand, if $n_l\neq 0$  one has a canonical commutation relation
that in turn implies that the involved operators
must  have {\it continuos spectrum}, 
which cannot hold in this case.
Notice also that equation (11) is wrong 
in that the exponentiated term $i\,n_l$  spoils unitarity.
At the end of a brief  calculation  the authors conclude 
that for the universal gate $n_l=0$, leaving  open the possibility
that other  values might be  allowed.
Moreover, such  calculation does not imply
the feasibility of universal gate operations
in that it simply reflects the fact that the encoded universal gate
 {\it by construction}  leaves the $X_{l_i},\,(i=1,2)$   eigenspaces
(that encode the information)
invariant and therefore commutes with $X_{l_i},\,(i=1,2).$\\
We conclude by observing that the proposed enconding
is nothing but 
the 'clusterized' form   of the one discussed in
 \cite{ZARA},\cite{ZANA}  as a very special case of 
the most general (collective) dissipation-decoherence
interaction there considered.
In such a general case to four physical bits are needed to encode one 
logical bit and no FHE is required.

\end{multicols}

\begin{references}
\bibitem{EFF}  Indeed the here proposed enconding
is nothing but the one introduced in \cite{PALMA}.
\bibitem{ZANA} The idea of using clusters of qubits collectively coupled with the environment
to encode stable quantum information is also discussed
in \cite{ZANA1} within a Master Equation formalism.
\end{references}
\end{document}